%% file: main.tex
\documentclass[sigconf]{acmart}

\usepackage{booktabs} 
\usepackage{caption}
\usepackage{subfig}
\usepackage{url}
\usepackage{epstopdf}
\usepackage{array}
\usepackage{pifont}
\usepackage{multirow}
\usepackage{amssymb}

\usepackage[latin1]{inputenc}

\copyrightyear{2017}
\acmYear{2017}
\setcopyright{rightsretained}
\acmConference{SIGIR 2017 Workshop on Neural Information Retrieval (Neu-IR'17)}{}{August 07--11, 2017, Shinjuku, Tokyo, Japan}
\acmPrice{}
\begin{document}

\title{Microblog Retrieval for Post-Disaster Relief: Applying and Comparing Neural IR Models}



\author{Prannay Khosla}
\affiliation{\institution{IIT Kanpur, India}}
\author{Moumita Basu}
\affiliation{\institution{IIM Calcutta, India; IIEST Shibpur, India}}
\author{Kripabandhu Ghosh}
\affiliation{\institution{IIT Kanpur, India}}
\author{Saptarshi Ghosh}
\affiliation{\institution{IIT Kharagpur, India; IIEST Shibpur, India}}


\begin{abstract}
\input{abstract}
\end{abstract}

%
%

\ccsdesc[500]{Information systems~Information retrieval}



\keywords{Microblog retrieval; Disaster; Neural IR; Word embeddings; Character embeddings}
\maketitle

\input{intro}

\input{related}

\input{dataset}
\input{baselines}

\input{neural}

\input{evaluation_identify}

\input{reuse}

\input{conclu}

\vspace{2mm}
\noindent {\bf Acknowledgement:} 
The authors thank the anonymous reviewers whose comments helped to improve the paper.
This research was partially supported by a
grant from the Information Technology Research Academy (ITRA),
DeITY, Government of India (Ref. No.: ITRA/15 (58)/Mobile/DISARM/05).
P. Khosla and S. Ghosh also acknowledge support from Microsoft Research India -- 
part of the work was done at the 2017 MSR India Summer Workshop on Artificial Social Intelligence.

\bibliographystyle{ACM-Reference-Format}
\bibliography{kripa_reference,twitter_reference,deep_learning_reference} 

\end{document}

%% file: abstract.tex
Microblogging sites like Twitter are important sources of real-time information on ongoing events, such as socio-political events, disaster events, and so on. Hence, reliable methodologies for microblog retrieval 
are needed for various applications. In this work, we experiment with microblog retrieval techniques for a particular application -- identifying tweets that inform about
resource needs and availabilities, for effective coordination of 
post-disaster relief operations. 
Traditionally, pattern matching techniques are adopted to identify such tweets. 
In this work, we experiment with a number of neural network based 
retrieval models,  including word-level embeddings and character-level embeddings,
for automatically identifying these tweets. 
We perform experiments over tweets posted during two recent disaster events, 
and show that neural IR models outperform 
the pattern-matching techniques of prior works.
We also propose two novel neural IR models which performs competitively with 
several state-of-the-art models.
Further, recognising that the large training time of neural IR models is an obstacle 
in deploying such models in practice, 
we also explore the reusability of neural IR models trained over past events, for
retrieval during future events.

%% file: intro.tex
\section{Introduction}

\noindent
Microblogging sites like Twitter and Weibo have emerged as important sources
of real-time information on ongoing events, including socio-political events, emergency events, and so on. 
For instance, during emergency events (such as earthquakes, floods, terror attacks),
microblogging sites are very useful for gathering
 situational information in real-time~\cite{social-media-emergency-survey,varga-help-tweets}. 
During such an event, typically only a small fraction of the microblogs (tweets) posted are relevant 
to the information need. Hence, it is necessary to design effective methodologies for microblog retrieval, so that
the relevant tweets can be automatically extracted from large sets of documents (tweets).  

Microblog retrieval is a challenging IR problem, primarily due to the noisy vocabulary and very short length of tweets. 
The 140-character limit on tweets prompts users to use arbitrary shortenings of words, and non-standard abbreviations~\cite{stemming-ecir17}. 
Additionally, different users can express the same information in very different ways. 
Hence, traditional Information Retrieval / Natural Language Processing techniques often
do not perform well on noisy microblogs. 
This limitation of standard methodologies has motivated the IR community in recent years to adopt
neural network based IR models~\cite{neural-models-ir-arxiv,neural-ir-lit-review-arxiv} for microblogs (see Section~\ref{sec:related}).

In this work, we apply and compare various neural network-based IR models for microblog retrieval for a specific 
application, as follows.
In a disaster situation, one of the primary and practical challenges in coordinating the post-disaster relief operations 
is to know about {\it what resources are needed} and 
{\it what resources are available} in the disaster-affected area. 
Thus, in this study, we focus on extracting these two specific types of microblogs or tweets.

\vspace{1mm}
\noindent {\bf Need-tweets:} Tweets which inform about the need or requirement
of some specific resources such as food, water, medical aid, shelter, etc.
Note that tweets which do not directly specify the need, but 
point to scarcity or non-availability of some resources are also
included in this category. 

\vspace{1mm}
\noindent {\bf Availability-tweets:} Tweets which inform about the availability
of some specific resources. This class includes both tweets which inform
about {\it potential availability}, such as resources being transported
or despatched to the disaster-struck area, as well as 
tweets informing about the {\it actual availability} 
in the disaster-struck area, such as food being distributed, etc. 

\begin{table*}[tb]
\centering
\small
\begin{tabular}{|p{0.9\textwidth}|}
\hline
\multicolumn{1}{|c|}{{\bf Need-tweets}}
\\ \hline
Mobile phones are not working, no electricity,  no water in \#Thamel, \#Nepal. \#earthquake  \#NepalQuakeRelief  \\ \hline
Over 1400 killed.  Many Trapped. Medical Supplies Requested. \\ \hline
@canvassss @skyasesh @YouthForBlood they are in search of blood donors for the people who are injured in earthquake... help \\ \hline
Nepalis,  r w/o water \& electricity. Water is essential to be supplied to the affected people in Nepal. \\ \hline
World Community \#Nepal needs humanitarian aid, rescue \& medical aid \#NepalEarthquake \\ \hline
\multicolumn{1}{|c|}{{\bf Availability-tweets}}
\\ \hline
\# Langar meals available at Sikh Gurdwara at Kupondol near Bagmati Bridge \#Nepal \#NepalQuakeRelief  \\ \hline
\#India to setup Field Hospital in \#Nepal by tomorrow morning to provide medical facilities \#NepalEarthquake  \\ \hline
4 PAF aircraft w/ rescue \& relief assistance, incl a 30-bed mobile hospital have left for \#Nepal  \\ \hline
Earthquake emergency numbers VIRed cross ambulance service Nepal +00977 422 8094  \\ \hline
can anyone we know pick the 2000 second hand tents from Sunauli and distribute it to the people in need in Nepal? \#NepalQuake
\\ \hline
\end{tabular}
\caption{Examples of need-tweets and availability-tweets posted during a recent disaster event (2015 Nepal earthquake).}
\label{tab:need-available-examples}
\vspace*{-5mm}
\end{table*}

\noindent Table~\ref{tab:need-available-examples} shows some examples of need-tweets and availability-tweets posted during the 2015 Nepal earthquake. 
Apart from the noisy vocabulary of the tweets (e.g., `without' abbreviated to `w/o', `including' abbreviated to `incl'), 
it can be observed that needs and availabilities are expressed in many diverse ways. While some tweets might be easier to retrieve
due to presence of intuitive terms like `need', `require', or `available', many of the tweets do {\it not} contain such terms.
Given the wide diversity in the tweets, use of neural IR models seems promising,
since they might be able to capture the semantic relationships among various terms.

\vspace{2mm}
\noindent {\bf Present work:} In this work, we apply and compare various neural network based  models for retrieval of need-tweets
and availability-tweets, including word-level embeddings (Word2vec~\cite{Mikolov2013}),
models that combine both word-level and character-level embeddings~\cite{IJCAI2015}, 
models using such combined embeddings with attention~\cite{CaoR16}, and so on. 
We also propose two novel models which combine word-level and character-level embeddings. 
We perform a comprehensive evaluation of the methodologies using 
tweets posted during two recent disaster events -- the Nepal earthquake in April 2015,
and the earthquake in Italy in August 2016. 
We observed that word-level embedding models usually perform better in terms of Recall,
while models combining word and character embeddings generally achieve better Precision. 
Further, the proposed models perform better than most of the state-of-the-art models.

Note that, traditionally, pattern matching based schemes have been employed by prior works
for identifying specific types of tweets~\cite{purohit-first-monday,emterms-iscram}.
We also compare the neural IR models  with pattern matching techniques
of prior works, and show that neural IR models are much more effective in microblog retrieval.

It can be noted that a primary obstacle in deploying neural IR models for retrieval during
ongoing events is the large time needed to train such models. 
To this end, we also explore the reusability of neural IR models trained over past events,
for retrieval during future events with minimal re-training. 

%% file: related.tex
\section{Related work} \label{sec:related}

\noindent  {\bf Application of neural network models over microblogs:}
Traditional IR / NLP approaches often do not perform well over microblogs, primarily 
due to their short size and noisy, informal vocabulary. 
As a result, neural network based IR models~\cite{neural-ir-lit-review-arxiv,neural-models-ir-arxiv,deep-learning-ir-sigir16}
are increasingly being applied over microblogs. 
For instance, Severyn {\it et al.} applied deep convolutional neural networks for sentiment analysis of tweets~\cite{tweet-sentiment-cnn}, while  
Wang {\it et al.} composing word embeddings with Long Short-Term Memory for identifying the polarity of tweets~\cite{polarity-prediction-lstm}.
Again, Ganguly {\it et al.} used neural IR models for retrieving code-mixed microblogs~\cite{code-mixed-microblogs},
while Ma {\it et al.} used recurrent neural networks for detecting rumors from microblogs~\cite{rumor-recurrent-neuralnet}.
Our prior work~\cite{stemming-ecir17} proposed a contextual stemming algorithm using word embeddings
for retrieving tweets posted during disasters. 

~\\
\noindent  {\bf Utilising online social media for disaster relief:}
In recent years, there has been a lot of work on utilizing Online Social Media (OSM) for aiding
disaster relief operations~\cite{social-media-emergency-survey}.
However, to our knowledge, there have been only a few prior works
that have specifically focused on 
the problem of identifying OSM posts that inform about 
need and availability of resources.
Varga {\it et al.}~\cite{varga-help-tweets} developed NLP techniques
to identify such tweets. However, a large fraction of the tweets in the dataset is in Japanese, and it is unclear whether the methodology in~\cite{varga-help-tweets}
can be readily applied to tweets in English.

Some prior studies also identified patterns / lexicons which
can be used to identify specific types of tweets, including tweets
informing about need and availability of resources~\cite{emterms-iscram,purohit-first-monday}.
To our knowledge, the most comprehensive set of such patterns 
has been proposed by Temnikova {\it et al.}~\cite{emterms-iscram}.
We observed that a large fraction of the patterns identified in this study (referred to as EMTerms),
can be used to identify need and availability of various types of resources. 

Thus, the task of identifying  need-tweets and availability-tweets
has traditionally been approached as a pattern matching task.
In the present work, we adopt a different approach -- we view the tasks as {\it Information Retrieval} (search) tasks
and use neural network based retrieval models for the tasks.
We demonstrate that neural IR methodologies perform better than the 
prior pattern matching approaches~\cite{emterms-iscram,purohit-first-monday} for this application.


%% file: dataset.tex
\section{Datasets} \label{sec:dataset}

\noindent This section describes the datasets used for the experiments in this work, and also 
how the gold standard for evaluating the methodologies was developed.

\subsection{Microblogs related to two disaster events}

\noindent For the present work, we collected tweets 
related to two major earthquakes that occurred in recent times --
(i)~the earthquake in Nepal and India in April 2015,\footnote{\url{https://en.wikipedia.org/wiki/April_2015_Nepal_earthquake}}
and 
(ii)~the earthquake in central Italy in August 2016.\footnote{\url{https://en.wikipedia.org/wiki/August_2016_Central_Italy_earthquake}}
For both the disaster events, we used the 
Twitter Search API\footnote{https://dev.twitter.com/rest/public/search}
to collect tweets that were posted during the days immediately following the event.
The queries `nepal quake' and `italy quake' respectively, were used
to collect the tweets relevant to the two events. 
In total, about 100K tweets were collected for the Nepal earthquake,
and about 180K tweets for the Italy earthquake. 
For this work, we consider only tweets in English, as identified by the Twitter language identification system.

It has been observed that tweets frequently contain duplicates and near-duplicates as the same information is often 
retweeted / re-posted by many users~\cite{Tao-duplicate-tweets}. 
Presence of duplicates can result in over-estimation of the performance 
of retrieval / extraction methodologies.
and can also create information overload for human 
annotators while developing the gold standard~\cite{TREC-microblog-2015}.
Therefore, we eliminated duplicate and near-duplicate tweets using a simplified version of the methodologies 
discussed in~\cite{Tao-duplicate-tweets}. 
Specifically, similarity of a pair of tweets was estimated
by the Jaccard similarity of the set (bag) of words contained in the two tweets (after ignoring stopwords, URLs and @user mentions).
If two tweets were found to be more similar than a threshold value, only one of the tweets was retained in the corpus.
 
After removing duplicates and near-duplicates, we obtained a set of 
{\it 50,068 tweets} for the Nepal earthquake dataset,
and {\it 70,487 tweets} for the Italy earthquake dataset.  
These sets were used for all experiments reported in this study.
For brevity, we will denote the two datasets as
{\it nepal-quake} and {\it italy-quake} respectively.

\subsection{Developing gold standards for evaluation}

\noindent Evaluation of the methodologies discussed
in this work required a gold standard containing the need-tweets
and availability-tweets contained in the datasets. 
We engaged three human annotators to develop this gold standard,  each of whom is proficient in English
and is a regular user of Twitter, but none of whom is an author of this paper. 
Each annotator was given the two datasets of tweets (nepal-quake or italy-quake), and was 
asked to identify all need-tweets and availability-tweets in both datasets.

Each annotator was first asked to identify need - tweets and availability - tweets {\it independently}, 
i.e., without consulting the other annotators.
While many tweets were identified by all three annotators in common, there were some tweets
which were identified by two or only one of the annotators.
Hence, we conducted a second phase, where all need-tweets and availability-tweets
that were identified by at least one annotator (in the first phase) were considered.
The gold standard set of need-tweets and availability-tweets were finalized through discussion with
all the annotators and mutual agreement.

Finally, through the human annotation process described above, the following number of tweets 
were identified -- {\it 499} need-tweets and {\it 1333}  availability-tweets for nepal-quake dataset, 
and {\it 177} need-tweets and {\it  233} availability-tweets for  the italy-quake dataset.
Note that, even though the italy-quake dataset is larger than the nepal-quake dataset, 
the italy-quake dataset has much fewer need-tweets and availability-tweets. 
Hence, retrieving these tweets is likely to be more difficult in case of the italy-quake dataset.

%% file: baselines.tex
\section{Baseline methodologies}  
\label{sec:baselines}
\noindent In this section, we discuss three baseline methodologies
for identifying need-tweets and availability-tweets. 

\subsection{Pattern matching baselines}   \label{sub:pattern-match-baselines}

\noindent As stated earlier, most prior studies have used pattern matching approaches for identifying specific types of tweets posted during disaster events, including need-tweets and availability-tweets. We consider two such studies as baselines, as described below.

\vspace{2mm}
\noindent (1) Purohit {\it et al.}~\cite{purohit-first-monday} proposed a set of 18 regular expressions
to identify tweets that ask for donation of resources, and tweets that inform about availability of resources to be donated. We obtained, on request, from the authors of~\cite{purohit-first-monday}, the 18 regular expressions and use 
these on our dataset to identify need-tweets and availability-tweets.

\vspace{2mm}
\noindent (2) Temnikova {\it et al.}~\cite{emterms-iscram} proposed a large set of patterns (referred
to as EMTerms) to identify specific types of tweets during emergencies. 
We employed three annotators (the same as those who developed our gold standard, as described 
in the previous section) to select those patterns
which are relevant to need and availability of resources.
The patterns in EMTerms are grouped into several categories, out of which
the annotators identified six categories as relevant to need and availability of resources.
These six categories contain {\it 953 patterns} in total.  
Table~\ref{tab:emterms} shows the six categories, along with some example
patterns in each category.

\begin{table}[tb]
	\centering
	\small
	\begin{tabular}{ | p{0.4\columnwidth} | p{0.1\columnwidth} | p{0.38\columnwidth} | }
		\hline
		{\bf Category Code and Name}  &{\bf \# Patterns} & {\bf Examples of patterns}\\
		\hline
		T06: Need of / offered supplies, such as food, water, clothing, medical supplies or blood & 297 & \{Number\} bags, aid, aids, bottled water, donate any supplies\\
		\hline
		T07: Volunteer or professional services needed or offered & 232 & volunteer heads,  relief aid, help victims\\ 
		\hline
		C02: Needs food, or able to provide food &40 & \{Number\} bags of rice, distributes food, donations like canned goods \\
		\hline
		C04: Logistics and transportation & 232 &  \{Number\} trucks, helicopter, rescue boats\\
		\hline
		C05: Need of shelters, including location and conditions of shelters and camps & 92 & \{Number\} homeless, camps, hotel, shelter, shelter kit \\
		\hline
		C06: Availability and access to water, sanitation, and hygiene & 59 & need clean water, no drinking water, restoring water \\
		\hline
	\end{tabular}
	\caption{Examples of patterns from EMTerms~\cite{emterms-iscram} that are related to need / availability of resources (as identified by annotators)}
	\label{tab:emterms}
\vspace*{-10mm}
\end{table}
\subsection{Language model baseline}   \label{sub:indri-baselines}

\noindent
We consider a language model-based IR methodology as a third baseline. 
Here  the need and availability of resources are considered as broad topics (information needs),
and tweets relevant to each topic are retrieved and ranked based on their
relevance to the topics.
We consider two stages in the retrieval process - first, an {\it initial query} is used
to retrieve tweets, and subsequently, the query is expanded by adding some terms
to the initial query, and another round of retrieval is performed with the expanded query.

\vspace{1mm}
\noindent \textbf{Pre-processing the tweets:} 
All tweets are pre-processed
by case-folding to lower case, removal of a standard set of English stopwords, URLs and user-mentions,
and subsequent stemming.

\vspace{1mm}
\noindent \textbf{Retrieval with initial query:} 
We start with initial queries consisting of a few terms selected based on our intuition and observation 
of need-tweets and availability-tweets in general.
For retrieval of need-tweets, we use an initial query consisting of two terms -- `need' and `requir' 
(which is the stemmed form of `require' or `required'). 
For retrieval of availability-tweets, we use the initial query consisting of three (stemmed) terms --
`avail', `distribut' and `send'.

We employ the Indri IR system~\cite{indri} for the retrieval.\footnote{\url{http://www.lemurproject.org/indri/}} 
The pre-processed tweets were indexed using Indri,  
and then ranked retrieval of tweets was done using the default language model based 
retrieval model of Indri~\cite{indri}. 

\vspace{1mm}
\noindent {\bf Query expansion:}
The motivation of the query expansion phase
is to add to the query, some dataset-specific (event-specific) terms, so that
more relevant tweets can be retrieved.
We apply the well-known Rocchio expansion scheme~\cite{Manning:2008} for determining 
the candidate expansion terms. 
After documents are retrieved using a particular (initial) query, the top-ranked $k$ (a small number) 
documents are assumed to be relevant, and certain terms are selected from the
top retrieved documents to expand the query.
Specifically, for each distinct term in the $k=10$ top-ranked tweets retrieved by the original query, we compute the
$tf \times idf$ Rocchio scores,
where $tf$ is the frequency of the term among the
10 top-ranked tweets, and $idf$ is the inverse document frequency of the term over the entire dataset.
The top $p = 3$ terms in the decreasing order of Rocchio scores are  
selected for expanding the query. 

\vspace{1mm}
\noindent We will compare the performance of these baselines with that of several neural IR
retrieval models described in the next section.


%% file: neural.tex
\section{Neural network methodologies}  \label{sec:neural-models}

\noindent We consider several types of neural network-based models for 
retrieving need-tweets and availability-tweets, as described below.

\subsection{Neural network models}

We  consider one word level embedding model, and four models which combine
word level embeddings and character level embeddings (out of which two are novel
models proposed in this work). 

~\\
\noindent {\bf (1) Word embeddings (W2V):} 
We use the popular Word2vec tool~\cite{Mikolov2013} as a representative word embedding model. 
We first train Word2vec on the tweets (of a certain dataset). 
Then we use word vector models to model the embedding of every token. 
We consider each tweet as a list of tokens $\{u_1, u_2, u_3, \ldots, u_n\}$ and therefore for every token we consider a window of size $k$. For example for token $u_i$ the window is $ \{u_{i-k},u_{i-k+1},\ldots, u_{i+k-1},u_{i+k}\}$. We look up the embedding of every token $u_i$ and try to predict from every token its context tokens. 
Embedding $u_i = W[u_i]$ where $W$ is $d_{wrd}$ (size of embedding) dimensional vector look up table for every token.
We try 
$$ out(token, \theta) = U(\theta)W[token] $$
where the $out$ function approximator computes the probability that a said token is in the context window of token under consideration. $\theta$ are the weights of the function approximator. 
$$ L_1(B ; \theta) = \frac{1}{B} \sum_{token \in B} out(token, \theta)\log{p^{wrd}_i}$$
where $p^{wrd}_i$ is the inferred probability that a word lies in the context of the token under consideration. \\
For training Word2vec, we use the skip-gram model, along with Hierarchical softmax. The hyerparameters embedding size $d_{wrd}$ was taken  to be $256$, context size was $5$, learning rate was $0.5$. We use Stochastic Gradient Descent for training. 

\vspace{2mm}
\noindent {\bf (2) Combining word embeddings with character-level embeddings:} 
We have used the following three models:

~\\
\noindent {\bf (i) WC:} This model, proposed in~\cite{IJCAI2015}, aims at inferring character level embedding along with word level embeddings, from word level context of the token under consideration. The model tries to generalize by inferring the token embedding from how the character occurs in its context.
The last layer of the model remains the same as W2V~\cite{Mikolov2013}, while the hidden layer combines word level and character level embedding before feeding into the last layer. The embeddings are trained to predict the context of the token under consideration, and therefore is predicted to encode the semantics of the language:
$$ E[u_i] = \lambda_2 W[u_i] + (1-\lambda_2)\frac{1}{N}\sum_{c_i \in u_i}^{N} C[c_i]$$
where $\lambda_2$ is a self learned parameter, $C$ is a $d_{chr}$ (size of embedding) dimensional vector look up table for every character in the character vocabulary, and $W$ is a $d_{wrd}$ dimensional vector look up table for every token in the vocabulary. 

The loss function is
$$ L(B ; \theta) = \frac{1}{B} \sum_{token \in B} out(token,\theta)\log{p^{tot}_i}$$ 
where $out(token,\theta) = U(\theta)E[u_i]$ where $p^{tot}_i$ is the inferred probability that the said token is in the context of the token under consideration. \\ 
The model is run with embedding size $d_{wrd} = d_{chr} = 256$, word level context size as $5$, and a learning rate of $0.5$ and Adam decay rate $\beta_1 = 0.001$. We use Adam Optimizer for training, it being more robust for deeper models. 

~\\
\noindent {\bf (ii) WCAL:} This model was originally proposed by Cao and Rei~\cite{CaoR16}. The main advantage of this model is that a {\it model which encodes memory}, can embed the morphological features of a word/token into its embedding and helps in predicting the context. 
Relating the morphological features of a word to its context, theoretically improves generalization of the model to out of vocabulary tokens, 
and also seemingly different tokens with the same morphological features will have similar embedding in the said model.

Therefore, in this model, we obtain character level embeddings and feed them to a biLSTM, and then apply an {\it attention layer} over the embedding before we combine them with word level embeddings. The biLSTM model is fed all the character embedding $C[c_1,c_2, \ldots, c_n]$ to give us $\{h^{f}_1,h^{f}_2, \ldots,h^{f}_n\}$ and $\{h^{b}_{n},h^{b}_{n-1}, \ldots, h^{b}_1\}$. We concatenate to get $h_i = [h^f_i,h^b_i]$ for every character. The embedding is
$$ E[u_i] = \lambda_2W[u_i] + (1-\lambda_2)\sum_{c_j \in u_i}\alpha_{i}(u_i)h_j $$ 
where $$\alpha_{j}(u_i) = \frac{exp(v^{T}tanh(W(\theta)c_j))}{\sum exp(v^{T}tanh(W(\theta)c_j))}$$
The softmax layer ensures $\Sigma \alpha_i(u_i) = 1$, which implies that $\alpha$ is effectively a probability distribution. $\lambda_2$ is a self learned parameter.

The loss function $$ L(B ; \theta) = \frac{1}{B} \sum_{token \in B} out(token,\theta)\log{p^{tot}_i}$$ 
where $out(token,\theta) = U(\theta)E[u_i]$ where $p^{tot}_i$ is the inferred probability that the said token is in the context of the token under consideration. \\
The model is run with embedding size $d_{wrd} = 256$ and $d_{chr} = 128$ and a learning rate of $0.5$ and Adam decay rate $\beta_1 = 0.001$. We use Adam Optimizer training.

~\\
\noindent {\bf (iii) WCA (proposed):} This is a novel scheme that we propose in this work. In this model, we try to encode the morphological features in the final embedding. To obtain these, we combine word level embeddings with character level embeddings after applying an attention layer over them. 
We try to give more importance to some sections of the token, and we learn what characters to give more importance in what configurations, 
by {\it learning the parameters of the attention layer}. 
This model is smaller than RNN or LSTM models for morphological features, 
and hence requires lesser time and data to train. We also expect it to work better over noisy data such as microblogs.

The model uses an attention layer that computes the values of attention for each character using the character embedding of every character in the token. 
The embedding is
$$ E[u_i] = \lambda_2W[u_i] + (1-\lambda_2)\sum_{c_j \in u_i}\alpha_{i}(u_i)C[c_j] $$ 
where $$\alpha_{j}(u_i) = \frac{exp(v^{T}tanh(W(\theta)c_j))}{\sum exp(v^{T}tanh(W(\theta)c_j))}$$
The softmax layer ensures $\Sigma \alpha_i(u_i) = 1$, which implies that $\alpha$ is effectively a probability distribution. $\lambda_2$ is a self learned parameter. 

The loss function is
$$ L(B ; \theta) = \frac{1}{B} \sum_{token \in B} out(token,\theta)\log{p^{tot}_i}$$ 
where $out(token,\theta) = U(\theta)E[u_i]$, while $p^{tot}_i$ is the inferred probability that the said token is in the context of the token under consideration. \\ 
The model is run with token or character embedding size as $d_{wrd} = d_{chr} = 256$ and a learning rate of $0.5$ and $0.005$ for word embedding and character embedding respectively. Adam decay rate $\beta_1$ is set as $0.001$. We use Adam Optimizer training.

~\\
\noindent {\bf (iv) WCInd (proposed):} 
This is another novel method that aims at inferring the semantics of a character inside a token, 
and hence tries to generalize the morphology of the tokens at a character level. 
We adopt the approach of predicting the tokens in the context of the said character, 
which would help us bring inferring intra-character features (e.g., the high frequency of \textit{q,u}) 
in vicinity of each other as a feature. 

We first obtain character level embeddings in the same way we obtained embeddings for words -- by training for context. 
For every token, we obtain the embedding by combining the embedding for tokens obtained as in the WC model (described above) 
with the embedding obtained by finding the mean of the character embeddings of the said token:
$$ E[u_i] = \lambda W[u_i] + (1-\lambda)\frac{1}{N}\sum_{c_i \in u_i}^{N} C[c_i]$$
where $C$ is a $d_{chr}$ dimensional vector look up table for every character in the character vocabulary, $W$ is a $d_{wrd}$ dimensional vector look up table for every token in the vocabulary, and $\lambda$ is a hyperparameter.

The loss function is
$$ L_2(B ; \theta) = \frac{1}{B} \sum_{c \in B} out(c,\theta)\log{p^{c}_i}$$ 
where $out(c,\theta) = U(\theta)C[w]$ while $p^{c}_i$ is the inferred probability that the said characer is in the context of the character under consideration. 

The embeddings for words are trained independently using Stochastic Gradient descent. The embedding size is $256$ and the learning rate is $1$, while the character embeddings of the same dimension, are trained using Stochastic Gradient Descent with learning rate as $0.05$. 
The hyperparameter $\lambda$ is set to $0.7$.


\subsection{Using the neural models for retrieval}

\noindent \textbf{Training the models:}
The models were trained using the hyperparameters as stated above. 
We trained the models for around $12$ epochs, on a single GPU. 
We found that the models generally give best results after training for around 8 epochs, 
except the biLSTM model (WCAL) which requires more training. 

\vspace{1mm}
\noindent \textbf{Pre-processing the tweets:} 
Similar to what was described in Section~\ref{sub:indri-baselines}, 
all tweets are pre-processed
by case-folding to lower case, removal of a standard set of English stopwords, URLs and user-mentions,
and stemming.

\vspace{1mm}
\noindent \textbf{Retrieval with initial query:} 
We use the same initial queries as described in Section~\ref{sub:indri-baselines} --
the terms `need' and `requir'  for retrieving need-tweets, and the terms 
`avail', `distribut' and `send' for retrieving availability-tweets.

For a particular query, we construct a {\it query-vector} by performing vector addition of the term-vectors of all terms in the query,
and then dividing the vector sum by the number of words in the query. 
Similarly, for each tweet (pre-processed), we construct a {\it tweet-vector} by adding the term-vectors of all terms 
contained in the tweet and then dividing the vector sum by the number of terms in the tweet. 
For retrieving tweets relevant to a query, we calculate the cosine similarity between the 
corresponding query-vector and each tweet-vector. We than rank the tweets in decreasing order
of the cosine similarity.
In mathematical terms, 
we find the list of tweets $T^{'}$ from the original list of tweets $T$ as  
$$ T^{'} = argsort_{t \in T}\,{cos(E[t],E[q])} $$ 
where $E[u]$ is the embedding of the set of tokens $u$.

\vspace{1mm}
\noindent {\bf Query expansion:}
We use an expansion technique that utilises the term embeddings learned by the neural network models.
For a particular neural model, we first retrieve tweets using the initial query, and consider the top $k = 10$ retrieved tweets.
To expand the initial query, we compute the cosine similarity of the query-vector (of the initial query) with the
term-vector of every distinct term (as learned by the neural model under consideration) in the top $k$ tweets.
We select those $p = 3$ terms for which
the term-vector has the highest cosine similarity with the query-vector.

Table~\ref{tab:query-expansion-nepal} states the query expansion terms identified by
the different neural models over the nepal-quake dataset.
It can be observed that different neural models identify widely different expansion terms for the same query.
Similar observations were made for the italy-quake dataset, which we omit for lack of space.

\begin{table}[tb]
\centering
\small
\begin{tabular}{|p{0.3\columnwidth}|p{0.5\columnwidth}|}
\hline
\textbf{Model} & \textbf{Expansion terms} 
\\  \hline
\multicolumn{2}{|p{0.9\columnwidth}|}{{\bf Need-tweets (original query: `need requir')}} 
\\ \hline
W2V~\cite{Mikolov2013} & \textit{help, ample, earthquake}
\\  \hline
WC~\cite{IJCAI2015}  & \textit{giv, water, must}
\\ \hline
WCAL~\cite{CaoR16} & \textit{nepal, india, victim}
\\  \hline
WCA (proposed) & \textit{petrol, unitedWithNepal, nepal}
\\  \hline
WCInd (proposed)  & \textit{nepal, ample, india}
\\ \hline  \hline
\multicolumn{2}{|p{0.9\columnwidth}|}{{\bf Availability-tweets (original query: `avail, send, distribut')}} 
\\  \hline
W2V~\cite{Mikolov2013}  & \textit{security, helpWithMsg, donate}
\\ \hline
WCAL~\cite{CaoR16} & \textit{acrossCountry, provide, givenepal}
\\  \hline
WC~\cite{IJCAI2015}  & \textit{nepalDisasterReliefByMsg, security, msgHelpEarthquakeVictims}
\\  \hline
WCA (proposed) & \textit{nepalDisasterReliefBymsg, earthqk, 2help}
\\  \hline
WCInd (proposed)  & \textit{unrepair, give, relief}
\\ \hline  
\end{tabular}
\caption{Query expansion terms obtained using different methodologies on the \protect\textbf{Nepal-quake dataset}}
\label{tab:query-expansion-nepal}
\vspace*{-4mm}
\end{table}

%% file: evaluation_identify.tex
\section{Evaluation of methodologies} \label{sec:evaluation}

\noindent 
We now evaluate the methodologies described in the previous sections,
by comparing the tweets retrieved by a methodology with the gold standard 
identified by human annotators (as described in Section~\ref{sec:dataset}).

\vspace{2mm}
\noindent \textbf{Evaluation measures:} In a disaster situation, 
it is important both to identify need-tweets and availability-tweets precisely (high precision),
as well as to identify as many of the need-tweets and availability-tweets as possible (high recall). 
Hence, we use the following evaluation measures --
(i)~Precision@100,  (ii)~Recall@1000, (iii)~F-score, and (iv)~MAP (overall).  

Note that the pattern matching methodologies (described in Section~\ref{sub:pattern-match-baselines}) identify {\it unordered sets} of tweets,
while the retrieval methodologies output {\it ranked lists} of tweets. 
We intend to compare all the methodologies in a common evaluation setting.
Hence, for the pattern matching methodologies, we consider all the matched tweets 
if the number of matched tweets is less than 1,000; 
otherwise, we randomly select a subset of 1,000 tweets out of the matched tweets, and measure Precision, Recall, and F-score.


\begin{table}[tb]
\centering
\small 
\begin{tabular}{|p{0.4\columnwidth}|c|c|c|c|}
\hline
   \textbf{Methodology}  & \textbf{Prec} & \textbf{Recall} & \textbf{F-score} & \textbf{MAP} \\ 
\hline
\multicolumn{5}{|c|}{{\bf Need-tweets}}\\
\hline
(Baseline) Patterns from~\cite{purohit-first-monday} 	&  0.008  & 0.054 & 0.015 & -- \\
\hline
(Baseline) EMTerms~\cite{emterms-iscram} 		&  0.03  & 0.058  & 0.029 & --  \\
\cline{2-5}
Random-1000 and Overall 						& 0.03 &  0.737 & 0.055 & -- \\
\hline 
(Baseline) Language model   					& 0.20  & 0.239  & 0.218 & 0.077 \\
\hline
(Baseline) Language model, Rocchio expansion   & 0.12  & 0.290  & 0.170 & 0.094 \\
\hline
\hline
W2V~\cite{Mikolov2013}  						& 0.22  & 0.397 & 0.283 & 0.145 \\
\hline
W2V~\cite{Mikolov2013} with expansion 		& 0.36  & {\bf 0.433} & 0.393 & 0.180 \\
\hline
WC~\cite{IJCAI2015}  							& 0.35  & 0.317  & 0.332 & 0.121 \\
\hline
WC~\cite{IJCAI2015} with expansion 			& 0.38 & 0.347 & 0.362 & 0.137 \\
\hline
WCAL~\cite{CaoR16} 							& 0.32 & 0.275 & 0.296 & 0.101 \\
\hline
WCAL~\cite{CaoR16} with expansion 			& 0.32 & 0.297 & 0.307 & 0.106 \\
\hline
WCA (proposed) 								& 0.56 & 0.388 & 0.458 & 0.201 \\
\hline
WCA (proposed) with expansion 				& {\bf 0.57} & 0.389 & {\bf 0.462} & {\bf 0.202} \\
\hline
WCInd (proposed) 							& 0.21 & 0.283 & 0.240 & 0.096 \\
\hline
WCInd (proposed) with expansion 				& 0.28 & 0.340 & 0.310 & 0.113 \\
\hline

\hline
\hline
\multicolumn{5}{|c|}{{\bf Availability-tweets}}\\
\hline
(Baseline) Patterns from~\cite{purohit-first-monday} 	&  0.005 & 0.012  & 0.007 & -- \\
\hline
(Baseline) EMTerms~\cite{emterms-iscram}  			& 0.064  & 0.047  & 0.054 & -- \\
\cline{2-5}
Random-1000 and Overall 						 	& 0.063 & 0.613 & 0.116 & -- \\
\hline
(Baseline) Language model    						& 0.230 & 0.268 & 0.247 & 0.139\\
\hline
(Baseline) Language model, Rocchio expansion 		& 0.230 & 0.268 & 0.247 & 0.139  \\
\hline
\hline
W2V~\cite{Mikolov2013} 					& 0.50 & {\bf 0.398 } & 0.427 & 0.336 \\
\hline
W2V~\cite{Mikolov2013} with expansion 	& 0.59 & 0.374 & 0.458 & {\bf 0.388} \\
\hline
WC~\cite{IJCAI2015} 						& 0.70 & 0.314  & 0.433 & 0.277  \\
\hline
WC~\cite{IJCAI2015} with expansion 		& 0.75 & 0.344 & 0.472 & 0.333 \\
\hline
WCAL~\cite{CaoR16} 						& 0.71 & 0.292 & 0.414 & 0.254 \\
\hline
WCAL~\cite{CaoR16} with expansion 		& 0.83 & 0.332 & 0.474 & 0.333 \\
\hline
WCA (proposed) 							& 0.84 & 0.344 & 0.488 & 0.334 \\
\hline
WCA (proposed) with expansion 			& {\bf 0.84} & 0.344 & {\bf 0.488 } & 0.335 \\
\hline
WCInd (proposed) 						& 0.61  & 0.260 & 0.365 & 0.229 \\
\hline
WCInd (proposed) with Expansion 			& 0.79 & 0.254 & 0.384 & 0.235 \\
\hline
\end{tabular}
\caption{{\bf Comparing methodologies for the \protect\textbf{Nepal-quake dataset}.}}
\label{tab:results-nepal}
\vspace*{-10mm}
\end{table}

\begin{table}[tb]
\centering
\small
\begin{tabular}{|p{0.4\columnwidth}|c|c|c|c|}
\hline
\textbf{Methodology}  & \textbf{Prec} & \textbf{Recall} & \textbf{F-score} & \textbf{MAP} \\ 
\hline
\multicolumn{5}{|c|}{{\bf Need-tweets}}\\
\hline
(Baseline) Patterns from~\cite{purohit-first-monday} 	& 0.003 & 0.091 & 0.006 & -- \\
\hline
(Baseline) EMTerms~\cite{emterms-iscram} 	& 0.013 & 0.073 & 0.022 & -- \\
\cline{2-5}
Random-1000 and Overall 							& 0.013 & 0.458 & 0.026 & -- \\
\hline
(Baseline) Language model   						& 0.04 & 0.158 & 0.063 & 0.007 \\
\hline
(Baseline) Language model, Rocchio expansion 		& 0.01 & 0.164 & 0.018 & 0.005 \\
\hline
\hline
W2V~\cite{Mikolov2013} 						& 0.05 & 0.18 & 0.078 & 0.024  \\
\hline
W2V~\cite{Mikolov2013} with expansion 		& 0.05 & {\bf 0.367} & 0.088 & {\bf 0.051} \\
\hline
WC~\cite{IJCAI2015}  								& 0.06  & 0.124  & 0.081 & 0.012 \\
\hline
WC~\cite{IJCAI2015} with expansion 				& 0.06 & 0.141 & 0.084 & 0.015 \\
\hline		
WCAL~\cite{CaoR16} 								& 0.05 & 0.102 & 0.067 & 0.010 \\
\hline
WCAL~\cite{CaoR16} with Expansion				& 0.07 & 0.158 & 0.097 & 0.019 \\
\hline
WCA (proposed) 									& 0.02 & 0.073 & 0.031 & 0.009\\
\hline
WCA (proposed) with expansion 					& 0.02 & 0.079 & 0.032 & 0.009 \\
\hline
WCInd (proposed) 								& 0.09 & 0.266 & 0.134 & 0.032\\
\hline
WCInd (proposed) with expansion 					& {\bf 0.10} & 0.271 & {\bf 0.146} & 0.035 \\

\hline
\hline
\multicolumn{5}{|c|}{{\bf Availability-tweets}}\\
\hline
(Baseline) Patterns from~\cite{purohit-first-monday} 	&  0.002 & 0.039 & 0.004 & -- \\
\hline
(Baseline) EMTerms~\cite{emterms-iscram}  	& 0.022 & 0.100 & 0.038  & -- \\
\cline{2-5}
Random-1000 and Overall 						 	& 0.023 & 0.575 & 0.043 & -- \\
\hline
(Baseline) Language model  						& 0.05 & 0.090 & 0.064 & 0.008 \\
\hline
(Baseline) Language model, Rocchio expansion 	& 0.04 & 0.103 & 0.058 & 0.005  \\
\hline
\hline
W2V~\cite{Mikolov2013}   						& 0.05 & 0.171 & 0.077 & 0.030  \\
\hline
W2V~\cite{Mikolov2013} with expansion  			& {\bf 0.12} & {\bf 0.335} & {\bf 0.176} & {\bf 0.060} \\
\hline
WC~\cite{IJCAI2015}								& 0.01 & 0.056 & 0.017 & 0.009 \\
\hline 
WC~\cite{IJCAI2015} with expansion				& 0.01 & 0.069 & 0.017 & 0.009 \\
\hline
WCAL~\cite{CaoR16} 								& 0.02 & 0.064 & 0.031 & 0.010 \\
\hline
WCAL~\cite{CaoR16} with Expansion 				& 0.04 & 0.056 & 0.046 & 0.006 \\
\hline
WCA (proposed)									& 0.01 & 0.030 & 0.015 & 0.008 \\
\hline
WCA (proposed) with Expansion 					& 0.01 & 0.030 & 0.016 & 0.007 \\ 
\hline
WCInd (proposed) 								& 0.03 	& 0.039 & 0.033 & 0.005 \\
\hline
WCInd (proposed) with expansion					& 0.03 & 0.05	& 0.033  & 0.007 \\

\hline
\hline
\end{tabular}
\caption{Comparing methodologies for the \protect\textbf{Italy-quake dataset}}
\label{tab:results-italy}
\vspace*{-10mm}
\end{table}

\vspace{2mm}
\noindent \textbf{Retrieval results:} Table~\ref{tab:results-nepal} shows the performance of 
various methodologies on the nepal-quake dataset,
while Table~\ref{tab:results-italy} shows the results on italy-quake dataset. 
It is evident that, across all methodologies, the performances are significantly better
over the nepal-quake dataset than over the italy-quake dataset, which again indicates
that need-tweets and availability-tweets are much more difficult to retrieve for the italy-quake dataset.\footnote{As stated in the section describing the
datasets, the italy-quake dataset is larger than the nepal-quake dataset, but contains much fewer need-tweets and available-tweets.}

The EMTerms~\cite{emterms-iscram} patterns (one of the baselines) matche a very large number
of tweets -- more than 12,000 for nepal-quake and more than 6,000 for italy-quake. 
Considering all the matched tweets, the overall recall achieved is the highest among all methods 
(e.g., $0.737$ for need-tweets and $0.613$ for availability-tweets in the nepal-quake dataset).
However, the matched tweets also include many non-relevant tweets, leading to very low precision values, and hence low F-scores.
In contrast, the retrieval methodologies achieve both reasonable precision as well as reasonable recall, 
leading to significantly better F-score values
than the pattern matching methods. 

The neural network based retrieval methodologies perform significantly better than both
the pattern matching techniques as well as the language model based IR techniques.
Among the different neural IR models, the word embedding models (W2V) usually achieve better Recall scores, while
the models combining word and character level embeddings achieve better Precision scores
(except in the case of availability-tweets in the italy-quake dataset).
Especially, the two proposed neural IR models perform better than most of the state-of-the-art models
in terms of MAP and F-score. 

Comparing the performance of retrieval with initial queries and that with expanded queries, we observe
that for almost all cases, query expansion helps to improve the performance. 
Hence, the embedding-based query expansion technique is effective in improving microblog retrieval.

%% file: reuse.tex
\section{Reusability of embeddings for retrieval during future events}

\begin{table*}[tb]
\centering
\small
\begin{tabular}{|p{0.4\columnwidth}||c|c|c|c||c|c|c|c|}
\hline
 & \multicolumn{4}{|c||}{{\bf After 1 epoch on nepal-quake}} & \multicolumn{4}{|c|}{{\bf After 5 epochs on nepal-quake}} \\
\hline
\textbf{Methodology}  & \textbf{Prec} & \textbf{Recall} & \textbf{F-score} & \textbf{Map} & \textbf{Prec} & \textbf{Recall} & \textbf{F-score} & \textbf{Map} \\ 
\hline
\multicolumn{9}{|c|}{{\bf Need-tweets}}\\
\hline
WCInd (proposed)							& 0.22 & 0.200 & 0.209 & 0.065 & 0.30 & 0.295 & 0.297 & 0.113 \\
\hline
WC~\cite{IJCAI2015}							& 0.34 & {\bf 0.315} & {\bf 0.326} & {\bf 0.120} & {\bf 0.37} & {\bf 0.319} & {\bf 0.342} & {\bf 0.127} \\
\hline
WCA (proposed)								& {\bf 0.35} & 0.259 & 0.297 & 0.105 & 0.35 & 0.283 & 0.313 & 0.119 \\
\hline
\hline
\multicolumn{9}{|c|}{{\bf Availability-tweets}}\\
\hline
WCInd (proposed) 								& {\bf 0.68} & 0.215 & 0.326 & 0.185 & 0.74 & 0.245 & 0.367 & 0.224 \\
\hline
WC~\cite{IJCAI2015}								& 0.66 & {\bf 0.236} & {\bf 0.347} & {\bf 0.215} & 0.67 & {\bf 0.309} & {\bf 0.423} & 0.294 \\
\hline
WCA (proposed) 									& 0.61 & 0.2251 & 0.329 & 0.195 & {\bf 0.75} & 0.293 & 0.421 & {\bf 0.269} \\
\hline
\end{tabular}
\caption{Using models pre-trained on \protect\textbf{italy-quake} dataset, for retrieval on \protect\textbf{nepal-quake} (after 1 and 5 epochs of re-training).}
\label{tab:reuse-italy-model-for-nepal}
\vspace*{-4mm}
\end{table*}

\begin{table*}[tb]
\centering
\small
\begin{tabular}{|p{0.4\columnwidth}||c|c|c|c||c|c|c|c|}
\hline
 & \multicolumn{4}{|c||}{{\bf After 1 epoch on italy-quake}} & \multicolumn{4}{|c|}{{\bf After 5 epochs on italy-quake}} \\
\hline
\textbf{Methodology}  & \textbf{Prec} & \textbf{Recall} & \textbf{F-score} & \textbf{Map} & \textbf{Prec} & \textbf{Recall} & \textbf{F-score} & \textbf{Map} \\ 
\hline
\multicolumn{9}{|c|}{{\bf Need-tweets}}\\
\hline
WCInd (proposed) 			& {\bf 0.08} & 0.129 & {\bf 0.099} & 0.015 & 0.09 & 0.124 & 0.104 & 0.019 \\
\hline
WC~\cite{IJCAI2015}			& 0.04 & 0.113 & 0.059 & 0.010 & {\bf 0.09} & 0.186 & {\bf 0.121} & {\bf 0.025} \\
\hline
WCA (proposed)				& 0.06 & {\bf 0.215} & 0.093 & {\bf 0.020} & 0.06 & {\bf 0.221} & 0.094 & 0.022 \\
\hline
\hline
\multicolumn{9}{|c|}{{\bf Availability-tweets}}\\
\hline
WCInd (proposed)			& 0.04 & 0.086 & 0.055 & 0.012 & 0.05 & 0.1 & 0.068 & 0.017 \\
\hline
WC~\cite{IJCAI2015}			& {\bf 0.12} & {\bf 0.197} & {\bf 0.149} & {\bf 0.044} & {\bf 0.12} & {\bf 0.253} & {\bf 0.183} & {\bf 0.055} \\
\hline
WCA (proposed) 				& 0.04 & 0.112 & 0.059 & 0.019 & 0.07 & 0.125 & 0.089 & 0.019 \\
\hline
\end{tabular}
\caption{Using models pre-trained on \protect\textbf{nepal-quake} dataset, for retrieval on \protect\textbf{italy-quake} (after 1 and 5 epochs of re-training).}
\label{tab:reuse-nepal-model-for-italy}
\vspace*{-5mm}
\end{table*}

A major problem in deploying neural IR models in practice, for tasks like microblog retrieval during an ongoing event, is the high training time for such models. 
For instance, most of the neural IR models described in the previous section required 4.5 -- 5 hours 
of training (over ~8 epochs) over the nepal-quake and italy-quake datasets, while the biLSTM model took even longer (and more epochs) to train.
During an ongoing event such as a disaster, when information needs to be retrieved quickly, 
it is often not practicable to allow such high training times.

One potential solution to this problem is to use {\it pre-trained models}, i.e., we pre-train the models
on one or more dataset(s) (e.g., tweets posted during past events) and then use the model for retrieval on
a new dataset (tweets posted during a future event) with minimal re-training.
In this section, we explore the possibility of reusing the neural IR models discussed earlier in the paper. 
Here we only experiment with the models which combined word and character embeddings. 
Also, for the experiments in this section, 
we only consider retrieval with the initial queries (without any query expansion).

In one set of experiments, 
we consider models pre-trained on the italy-quake dataset, and re-train the models over the 
nepal-quake dataset for (i)~{\it just one epoch}, and (ii)~five epochs.
Table~\ref{tab:reuse-italy-model-for-nepal} shows the retrieval performance of such 
models over the nepal-quake dataset.
Similarly, we took the models pre-trained over the nepal-quake dataset, and 
trained them over the italy-quake dataset for one / five epochs; Table~\ref{tab:reuse-nepal-model-for-italy} shows
the performance of such models over the italy-quake dataset.

As expected, we observe a trade-off between the training time (number of epochs trained) and the 
retrieval performance. 
For instance, the best MAP-score achieved for the nepal-quake need-tweets
was $0.201$ (from Table~\ref{tab:results-nepal}, without considering query expansion), 
obtained via full training over the nepal-quake dataset (which needed close to 4.5 hours).
On the other hand, the models pre-trained over italy-quake achieved MAP of $0.120$ after just a single epoch 
re-training on the nepal-quake data (Table~\ref{tab:reuse-italy-model-for-nepal}), which needed less than 30 minutes of re-training.

The performance of pre-trained models is even better over the italy-quake dataset where retrieval is more challenging
(as indicated in the previous section).
For instance, the models pre-trained on nepal-quake dataset achieves a MAP score of $0.044$
for the italy-quake availability-tweets after only a single epoch of re-training, which is 
higher than the MAP score achieved by any of the models when trained fully on the italy-quake dataset (Table~\ref{tab:results-italy}, without considering query expansion).
In fact, most of the models give better performance for the italy-quake dataset
when pre-trained over the nepal-quake dataset, than when trained only over the italy-quake dataset.

These experiments point out the potential for reusing models pre-trained over past events
for retrieval during future events, with minimal retraining.
Especially, if the new datasets have very sparse relevant information (as is the case for the italy-quake dataset), then 
pre-training on prior datasets can be helpful in improving retrieval performance, along with minimising re-training time.
In future, we look to explore the reusability of models pre-trained over several events together,
for retrieval during future events.

%% file: conclu.tex
\section{Conclusion}  \label{sec:conclusion}

\noindent 
We compared  different methodologies for retrieving two specific types of tweets / microblogs 
that are practically important for post-disaster relief operations, viz., need-tweets and availability-tweets. 
Using datasets of microblogs posted during two recent disaster events, we compared among pattern matching techniques,
language model based techniques, and several neural IR models for the same task.
We also proposed two neural IR models that combines word-level and character-level embeddings,
and performs competitively with several state-of-the-art models for the said microblog retrieval problem. 
We also explored the possibility of reusing neural IR models pre-trained over past events,
with the objective of minimising training time over new datasets. 

In future, we plan to experiment with neural IR models for microblog retrieval from other
standard datasets (e.g., the TREC microblog datasets). Also, we plan to further explore
the possibility of reusing pre-trained neural IR models for practical tasks such as retrieval during 
disasters, which might facilitate deployment of such models over fast changing data streams.